\documentclass[11pt,a4paper]{article}
\usepackage[utf8]{inputenc}
\usepackage[T1]{fontenc}
\usepackage[english]{babel}
\usepackage{amsmath,amsthm}
\usepackage{amsfonts}
\usepackage{amssymb}
\usepackage{graphicx}
\usepackage{subfig}
\usepackage[left=2cm,right=2cm,top=2cm,bottom=2cm]{geometry}
\usepackage{authblk}
\usepackage{xcolor}

\title{Space Radiation Hazards and Future Mars Missions\footnote{Confidential Manuscript and submitted to Heliyon}}

\author[$\dagger1$]{Kamsali Nagaraja}
\author[]{S. C. Chakravarty}
\author[]{Praveen Kumar Basuvaraj}

\affil[]{Department of Physics, Bangalore University, Bengaluru 560056 India}
\affil[$\dagger$]{Corresponding author: kamsalinagaraj@bub.ernet.in}
\date{}

\begin{document}
\maketitle

\begin{abstract}
Manned Mars missions planned in the near future are of very low solar activity period and hence higher than acceptable radiation doses mainly due to the Galactic Cosmic Rays (GCR) that would require special techniques and technological development for maintaining the good health of the astronauts. The present study is to make an assessment and characterise the coming years in terms of solar activity and space radiation environment especially due to the abundance of highly energetic heavy-ions also known as HZE charged particles. These particle fluxes constitute a major hazard to the astronauts and also to the critical electronic components of the spacecraft. Recent data on the HZE species (from $B$ to $Ni$) obtained from Advanced Composition Explorer (ACE) spacecraft shows a clear enhancement of the particle fluxes between the solar cycle 23 and solar cycle 24 (between sunspot number peaks of 2002 and 2014) due to the persisting low sunspot numbers of the latter cycle. The peak values of these cosmic ray fluxes occur with a time lag of about a year of the corresponding minimum value of the sunspots of a particular 11-year cycle which is pseudo-periodic in nature. This is demonstrated by the Fourier and Wavelet transform analyses of the long duration (1700-2018) yearly mean sunspot number data. The same time series data is also used to train a Hybrid Regression Neural Network (HRNN) model to generate the predicted yearly mean sunspot numbers for the solar cycle 25 ($\approx$ 2019-2031). The wavelet analysis of this new series of annual sunspot numbers including the predictions up to the end of 2031 shows a continuation of the low solar activity trend and hence continuation of very high HZE fluxes prevailing in solar cycle 24 into the solar cycle 25 and perhaps beyond.
\end{abstract}


\section{Introduction}

Astronauts undertaking future space voyages to Mars would need to expend longer periods of at least 1~to~2 years to gather information about its surface features, geological, environmental and atmospheric phenomena, including efforts for a confirmatory evidence of possible existence of life, either in the present or the past. Extending this goal to gear up for a human settlement in Mars, plans are afoot to realise manned Mars missions within the next decade (NASA, 2013). In addition to general safety considerations, mission to Mars involving such long duration stays would need to be carefully designed to minimise the doses of potentially damaging solar and cosmic radiations which could constrain the planetary work schedule to the extent of jeopardising the mission objectives. While in transit through Earth’s magnetosphere the spacecraft shielding brings down the radiation effects considerably, the situation in Martian space would be harsher with only relatively marginal natural shielding of the atmosphere and the non-existent magnetosphere (ICRP, 2013). 

The radiation exposure measured by the absorbed dose, $D$, is the mean energy deposited per unit mass:

\begin{equation}
D=\frac{d\overline{E}}{dm}
\end{equation}

where $\overline{E}$ is the mean energy and $m$ the mass of the absorbing matter, $D$ is given in units of Gray (Gy) and 1 Gy is equal to 1 J kg$^{-1}$ which is equal to 100 rads, one rad being equivalent to the absorption of 100 ergs per gram.

For biological tissues, each type of radiation has an ionisation potential, determining the equivalent dose ($H_T$):  
\begin{equation}
H_T=\sum_{R}W_R \times D_{T,R}
\end{equation}  

where $D_{T,R}$ is the absorbed dose for radiation type $R$; $T$ stands for a particular tissue and varying with the type of radiation, is the quality or weighting factor. 

The equivalent dose is measured in Sieverts (Sv), which is 1 J~kg$^{-1}$, equal to 100 Roentgen (ICRP,~2007). To illustrate the potential risks involved, the Apollo astronauts received average radiation doses of 1.6 mGy (1~mGy = 0.001~Gy) to 14~mGy over two weeks. Assuming a quality factor of around 4, the missions resulted in an equivalent dose of 6.4~mSv to 56~mSv during the return Moon trips, which is greater than the allowable radiation workers’ yearly dose limit of 50~mSv (5~rem) and maximum public allowable exposure limit of 1~mSv (0.1~rem) per annum. The major components of space radiation consist of high energy ionising charged particles such as heavy ions (from $Be$ to $Ni$), protons and beta particles. Through  nuclear interactions of these charged particles with  materials of spacecraft, planetary surface, atmosphere, base structures, and the space suits of astronauts, secondary radiation of energetic albedo neutrons and protons can be produced further enhancing the overall radiation exposure. For a long duration manned Mars mission, the high energy (primarily in the hundreds of MeV to many GeV range peaking around 1~GeV) protons and high atomic number ions called HZE particles of the Galactic Cosmic Rays (GCR) are the major contributors to space radiation doses (Simpson et al., 1983; O’Neill, 2006). Due to their high energies and high Linear Energy Transfer ($dE/dx$) values, it is difficult to shield HZE radiations and make a realistic estimate of experimentally verified biologically effective doses. High energy (in keV-MeV range) solar charged particle radiations with a component of moderate energy and high $Z$ ions are also important to consider. However the exposure to these radiations could be controlled by nominal shielding and avoiding open space ventures during intense eruptive processes of the Sun called Solar Particle Events (SPE) as these can be predicted to a degree of reliability (Bertsch et al., 1969; Kim et al., 2006). 

The intensity of the GCR flux varies over the 11-year solar cycle due to the changes in the interplanetary plasma emanating from the expanding solar corona resulting in maximum doses during the solar minimum year. The equivalent annual radiation dose from GCR in interplanetary free space has been estimated to be about 0.73 Sv year$^{-1}$ and 0.28 Sv year$^{-1}$ during solar minimum and solar maximum years respectively. While the dose rates (estimated using OLTARIS website vary between 0.33 and 0.08 Sv year$^{-1}$ on the surface of Mars due to planet self-shielding and some attenuation through the thin Martian $CO_2$ atmosphere of about 16 g~cm$^{-2}$, using further aluminium shielding of even 20 g~cm$^{-2}$ does not reduce the dose rates appreciably (0.26 \& 0.07 as against 0.33 \& 0.08 Sv year$^{-1}$). Hence the prevention against the GCR radiation doses is the main challenge to undertake manned Mars missions particularly during the solar minimum period. The main purpose of the present research is to examine the details of the variations of solar activity during solar cycles 23, 24 and 25 covering the period up to $\approx$ 2031 and understand the limits of risks involved in carrying out manned Mars missions in the near future. Suitably shielded robotic Mars mission however, may prove to be safer even during the low solar activity conditions with high levels of radiation environment.

\section{Observational Data and Method of Analysis}
As mentioned above, future Mars missions would need to guard from the deleterious space radiation primarily due to the continuous influx of GCR charged particles which are difficult to filter out by reasonable shielding techniques on Martian surface. As already seen the temporal variation of this radiation is anti-correlated with the changes in solar activity revealed mainly by the sunspot numbers. However in the present context, it is necessary to have a better understanding and a quantitative assessment of the variation of long term changes in solar activity with its inherent periodicities and possible predictions up to the year 2031 or till the end of solar cycle 25. The results of this analysis along with the recent data on GCR fluxes from spacecraft measurements are examined for projecting the levels of this radiation in the years to come.  

The internationally revised daily and annual mean sunspot number (SSN) data used here are obtained from the World Data Centre (WDC) for the production, preservation and dissemination of the international sunspot number in Brussels (http://www.sidc.be/silso). The GCR data has been downloaded from Advanced Composition Explorer (ACE) Science Centre covering the period 1977 to 2019.

Hybrid Regression Neural Network (HRNN) technique is used with a programme code generated by Daniel Okoh and available to users (https://in.mathworks.com/matlabcentral/fileexchange/65686) to predict the smoothed yearly mean SSN for the solar cycle 25. The SSN time series with and without these predictions are subjected to Fast Fourier Transform (FFT) and Wavelet Transform analyses to determine the relative amplitudes of the main periodicities and their combined effect on variation of SSN in the coming decades, as this is one of the key parameters to predict GCR radiation environment in space and on Mars.

\section{Results and Discussion}
The charged particle fluxes of GCR are measured by the Cosmic Ray Isotope Spectrometer (CRIS) payload on board the Advanced Composition Explorer (ACE) spacecraft that was launched around $L_1$ point on August 25, 1997 (Stone et al., 1998). It is designed to measure the elemental and isotopic composition of GCR over 7-energy bands spanning $\approx$ 50-500 MeV per nucleon. The energy bands are different for each element. The energy bands are provided along with the fluxes/counts of the ionic species by the ACE Science Centre (http://www.srl.caltech.edu/ACE/ASC/index.html).

\begin{figure}[h]
	\vspace*{0cm}
	\centering
	\makebox[0pt]{%
	\includegraphics[height=10.5cm,width=0.8\paperwidth]{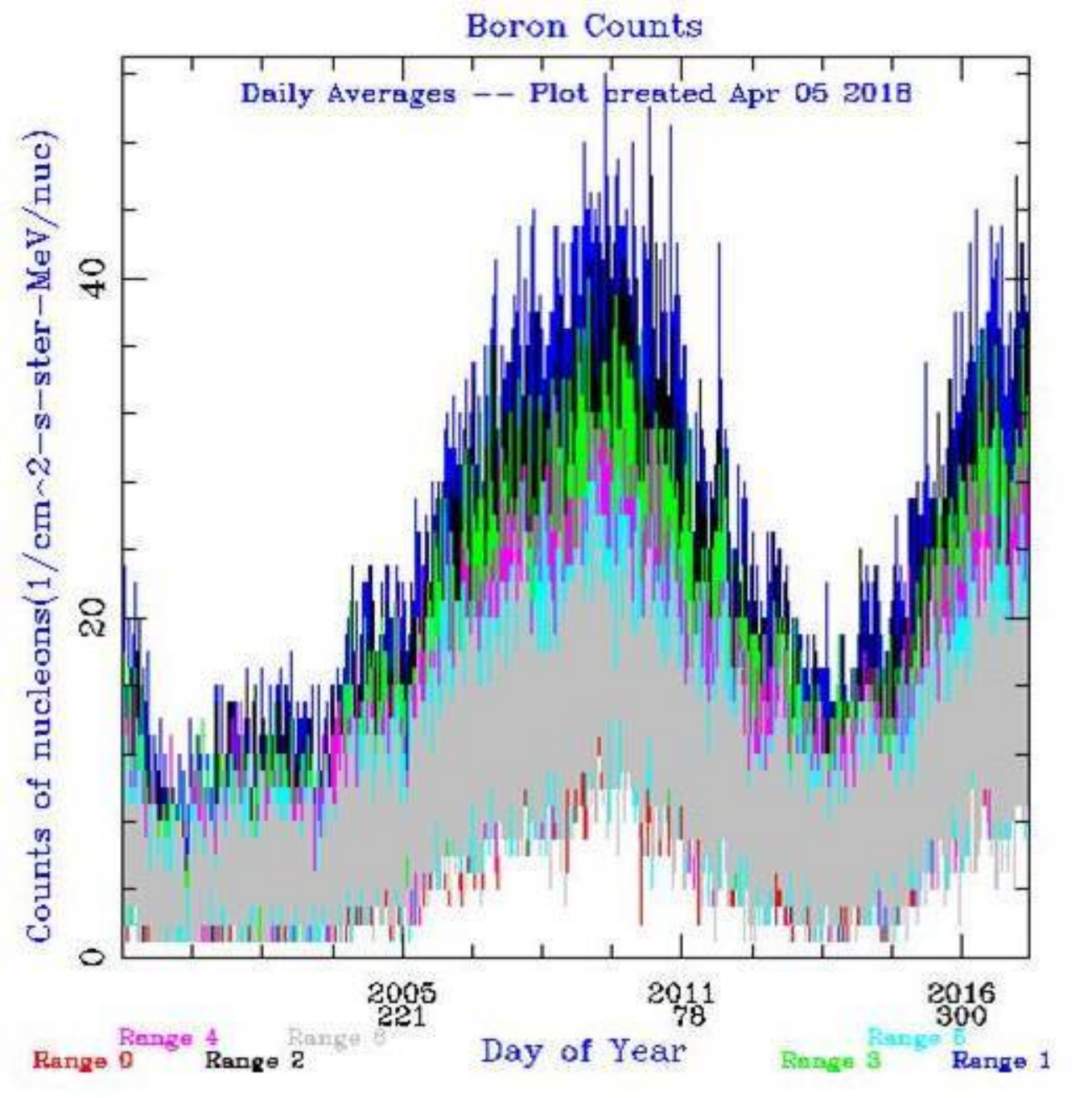}}
	\caption{CRIS-ACE Level-2 Data plot of Boron ion counts during 1~Jan~2000 to 2~Mar~2018 for 7 different energy range between 51.4 and 160.2 MeV per nucleon.\label{overflow}}
\end{figure}

Figure 1 shows a sample plot (facilitated by the data centre) for daily average Boron ion counts in 7 energy bands from 51.4 to 160.2 MeV per nucleon during the period 1~Jan~2000 to 2~Mar~2018 covering two solar minimum (2009, 2018-19) and two solar maximum (2002, 2014) activity years. While the inherent inaccuracy in the measurements is not a limiting factor to discern relative temporal variabilities between different energy bands, detailed information regarding the calculation of CRIS intensities and related errors are given in the references (George et al., 2009; Lave, 2012a; 2012b). The energy band associated colour-coded particle counts show decreasing trend with increasing energy of Boron ion and its anti-correlation with solar activity for all energy bands; counts peaking near the minimum solar activity years. Similar characteristic variations are observed in other 23 plots (not shown here) pertaining to the elements from Carbon ($C$) to Nickel ($Ni$). 

\begin{figure}[h]
	\vspace*{0cm}
	\centering
	\makebox[0pt]{%
	\includegraphics[height=10.5cm,width=0.8\paperwidth]{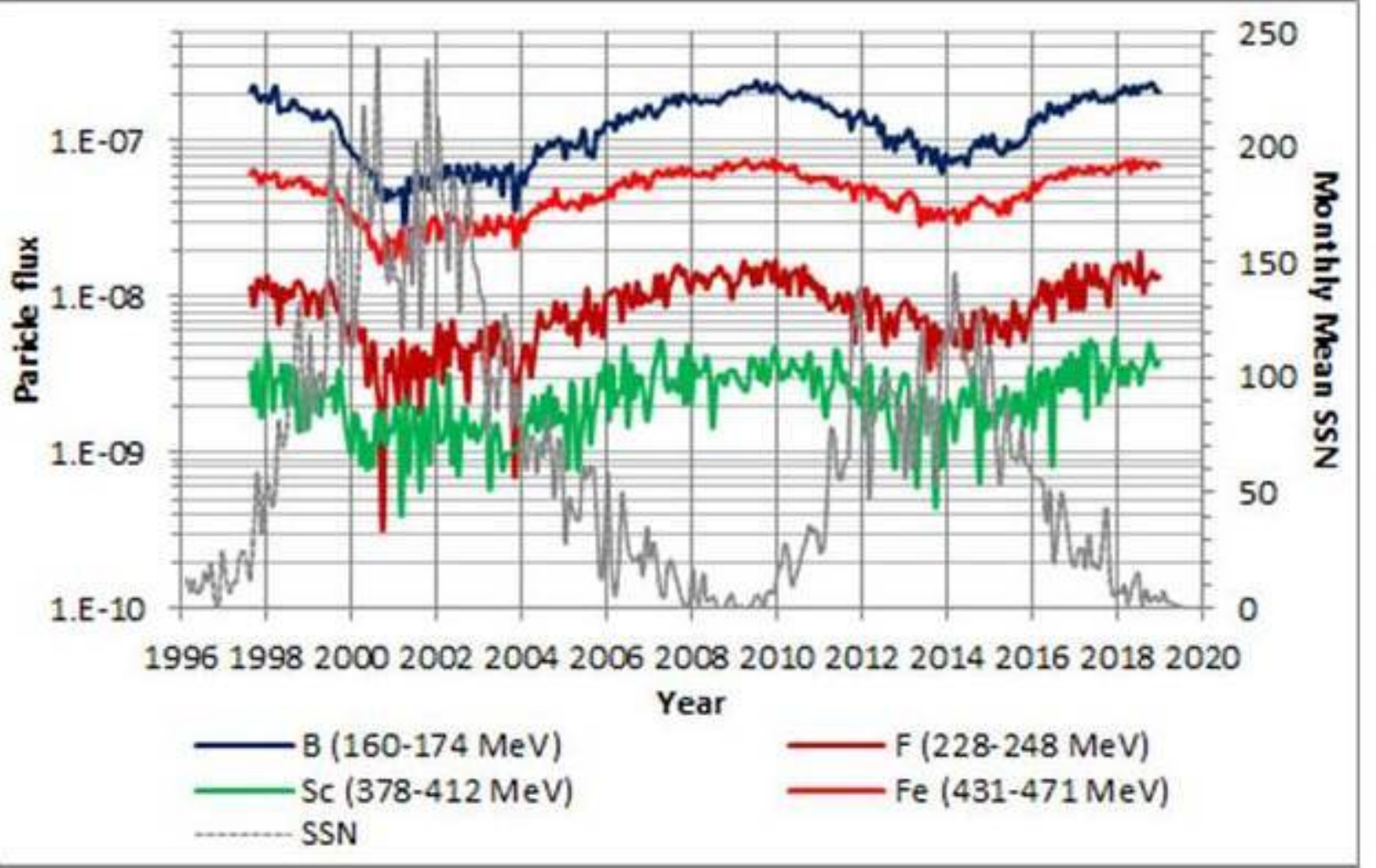}}
	\caption{Data plots of Bartels rotation average cosmic ray element fluxes of Boron ($B$), Fluorine ($F$), Scandium ($Sc$) and Iron ($Fe$) ions during 1997 to 2018, for roughly similar energy bands. The monthly mean SSN are also shown for comparison.}
\end{figure}

The level 2 CRIS data is organised into 27-day time periods i.e., roughly one solar rotation period and also known as Bartels rotation. For each Bartels rotation time average particle flux values are provided for 24 elements ($C$ to $Ni$), in units of particles/(cm$^{2}$~sr~sec~MeV/nucleon), in 7 energy bands mentioned above. Figure 2 shows sample time series plots of selected ion particle fluxes of $B$, $F$, $Sc$ and $Fe$ of atomic numbers 5, 9, 21 and 26, which are representative of the general pattern of long term variation with solar activity. The monthly average sunspot numbers for the solar cycles 23 and 24 are also plotted as secondary Y-axis. 

\begin{figure}[h]
	\vspace*{0cm}
	\centering
	\makebox[0pt]{%
	\includegraphics[height=9.5cm,width=0.8\paperwidth]{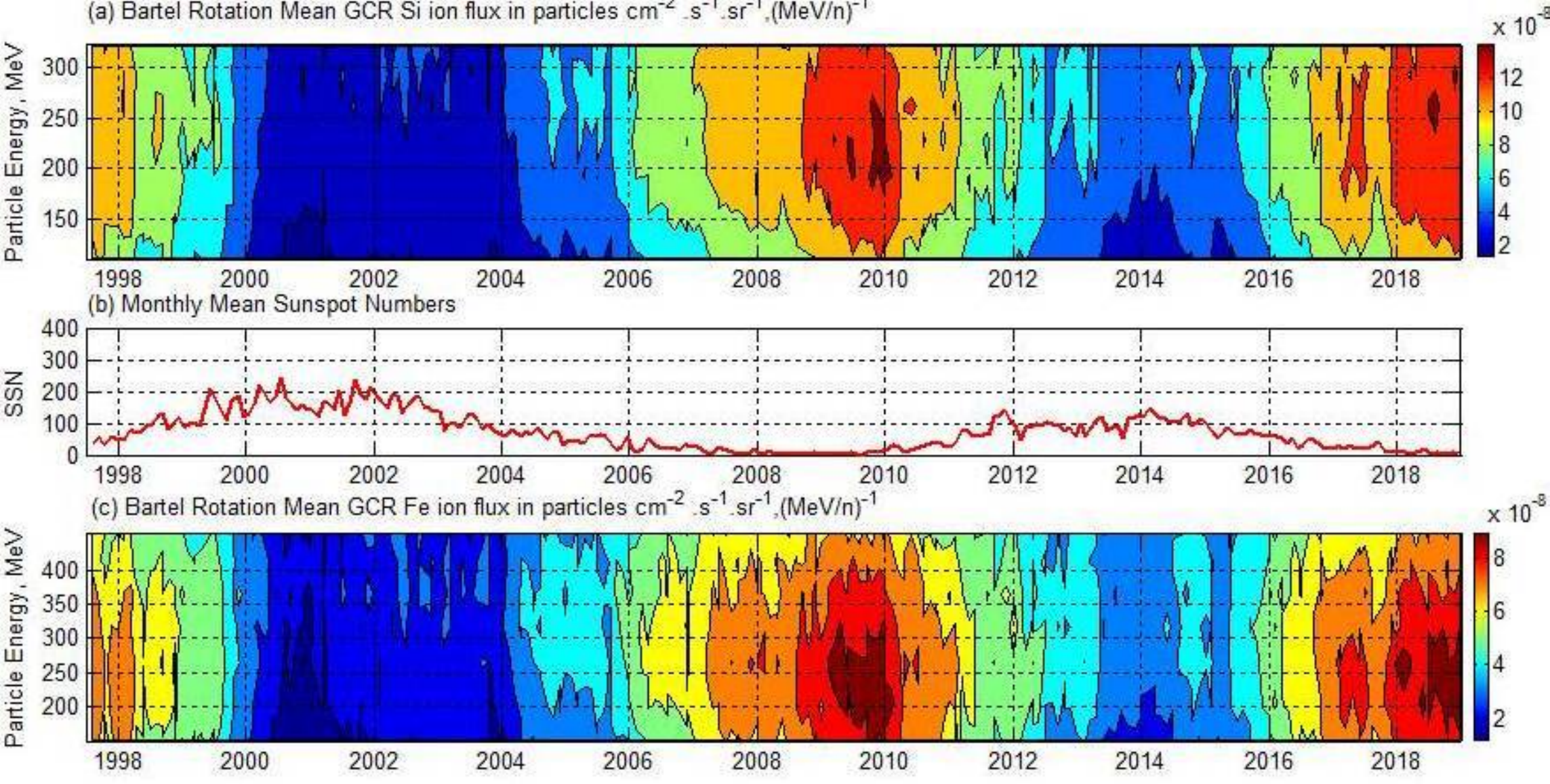}}
	\caption{Average cosmic ray element fluxes of Silicon ($Si$) and Iron ($Fe$) ions during 1997 to 2018 for Bartels rotations, covering observations in all the energy bands along with the monthly mean sunspot numbers.}
\end{figure}

The anti-correlation between the GCR flux and SSN can be seen for all the 4-GCR elements. Flux values of GCR elements $B$ and $Fe$ are higher by more than one order of magnitude compared to those of $F$ and $Sc$ throughout the observation period. While the GCR flux values are considerably lower at SSN peak years (2002 and 2014), the overall level of fluxes is considerably higher during the period of solar cycle 24 as compared to solar cycle 23 since the Sun is going through a very low 11-year activity phase under solar cycle 24. This result is further expanded in Figure 3 showing contours of particle fluxes ($Si$ and $Fe$) using the same data set but retaining the information of the full energy coverage of all the 7-bands. The plots indicate the following characteristics related mainly to solar cycles 23 and 24: (a) the flux intensities peak around 200-300 MeV of particle energies, (b) lower SSN values in cycle 24 has given a prolonged period of higher GCR fluxes between 2007 to 2018 when the 11-year SSN values were very low compared to the previous 11-years and, (c) there is a lag of at least one year between the SSN minimum and flux maximum evidenced from the concurrent plot of monthly mean SSN values. The figure also provides an unconfirmed clue that the low solar activity related GCR fluxes would continue to be higher by a factor of 2 in the coming years if the trend of low solar activity persists for the next cycle 25. The trends also suggest that the starting of cycle 25 may be delayed or may be slow to pick up in solar activity resulting in continuation of enhanced GCR fluxes.

\begin{figure}[h]
	\vspace*{0cm}
	\centering
	\makebox[0pt]{%
	\includegraphics[height=6.5cm,width=0.8\paperwidth]{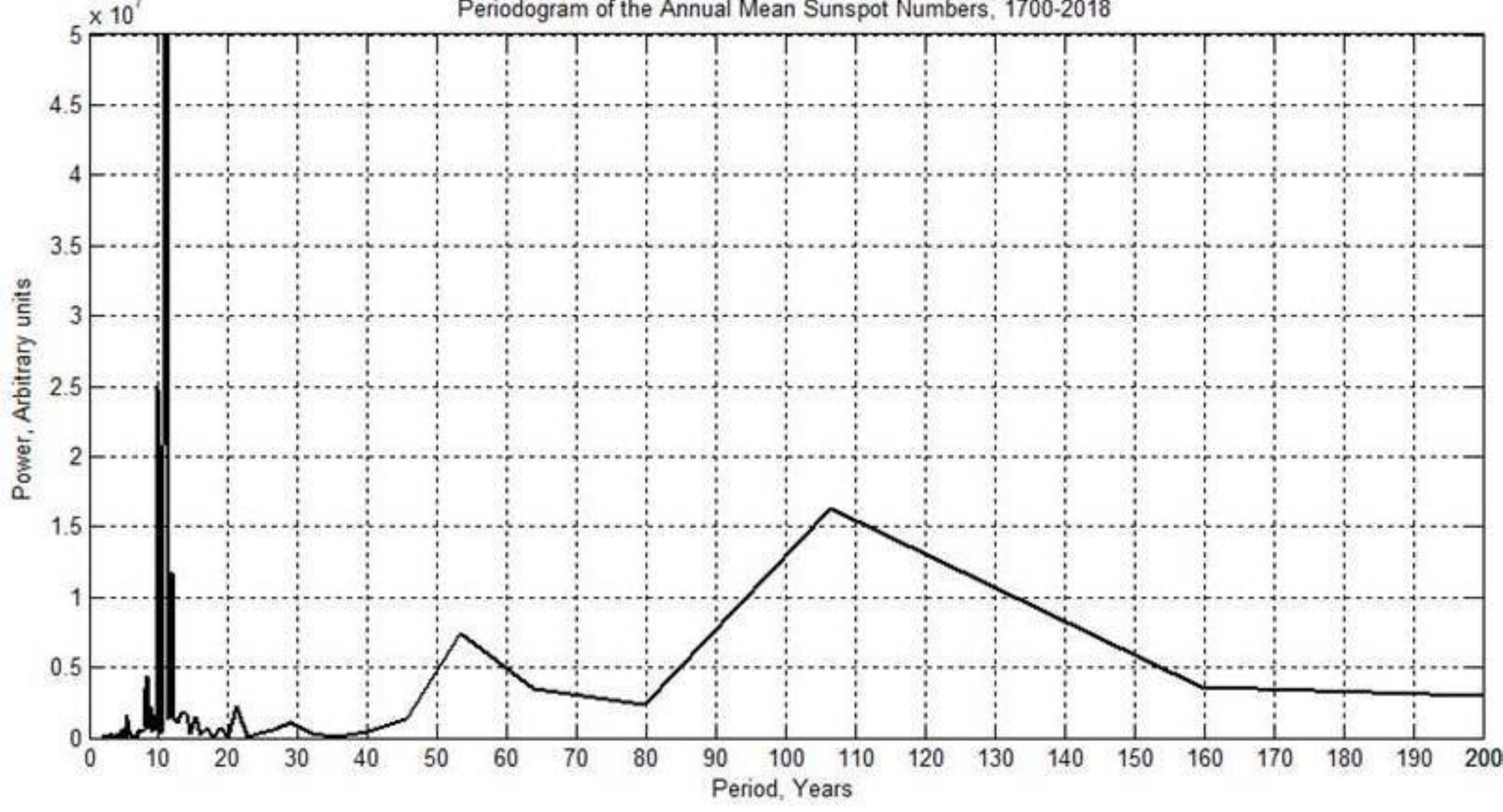}}
	\caption{Power spectrum of yearly mean SSN data series (1700 to 2018) showing discrete frequencies/periods of prominent solar activity cycles of approximately 11, 22, 53 and 107 years.}
\end{figure}

Results presented so far and further studies in this direction require a better appreciation of the long term variation of solar activity which is critical to make a more quantitative assessment of the radiation environment in free space or on Martian surface for possible manned Mars missions in the near future. In view of this and taking advantage of  the availability of a very long series of SSN data from 1700, a detailed analysis has been carried out using FFT, Wavelet and Neural Network techniques. The results provide some insights into the variation of the periodicities of solar activity so far, as well as into the future particularly during the solar cycle 25. Figure 4 shows the FFT frequency spectrum of the yearly mean sunspot numbers for the time interval 1700 to 2018. The discrete frequencies are equivalent to solar cycles of 11, 22, 53 and 107 years. It is seen that 11-year periodicity has maximum power followed by that of 107- (about 20\% power compared to the 11-year cycle), 53- and 22-year periodicities. 

\begin{figure}[h]
	\vspace*{0cm}
	\centering
	\makebox[0pt]{%
	\includegraphics[height=10cm,width=0.8\paperwidth]{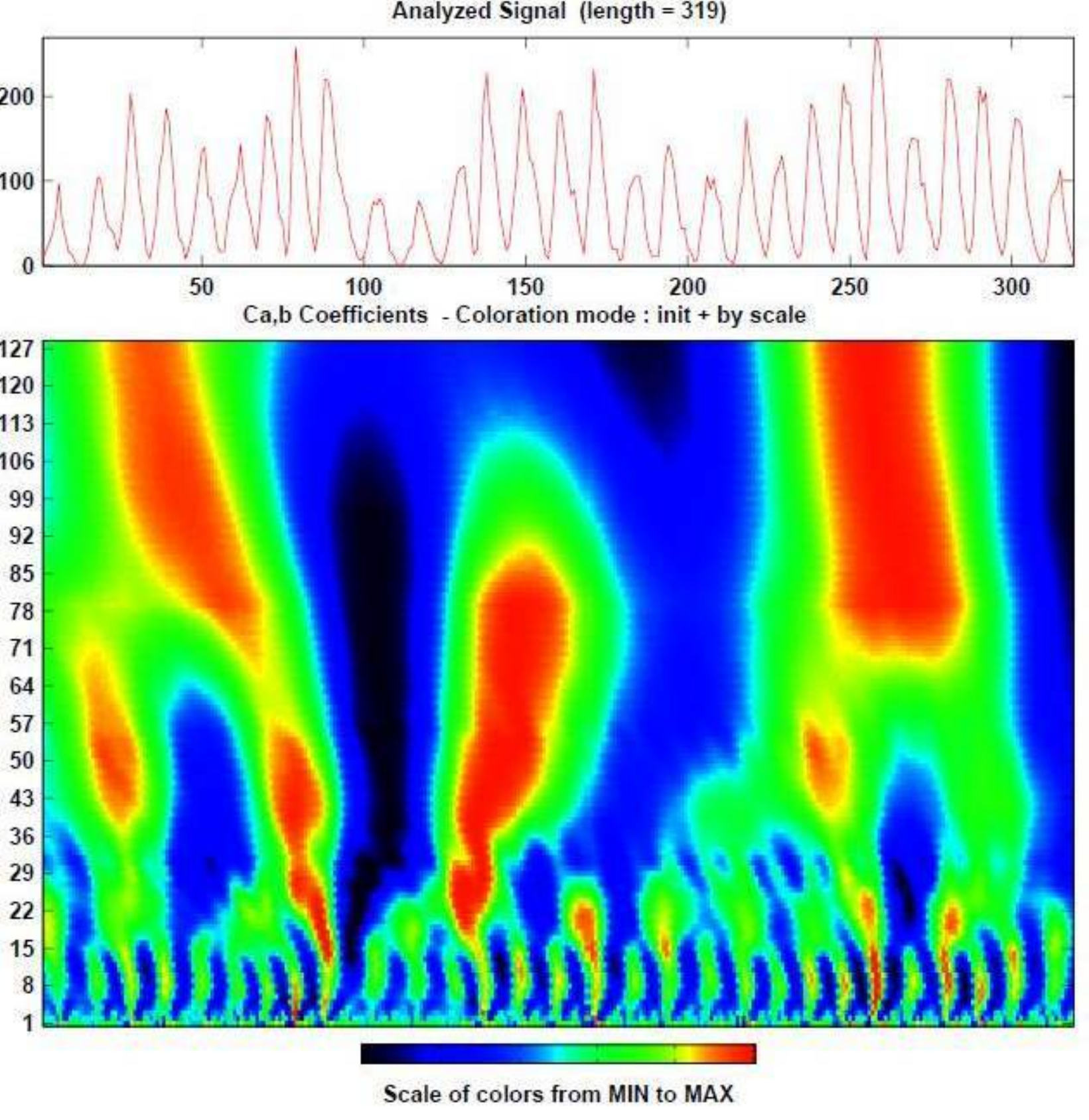}}
	\caption{Top: Yearly mean SSN (Y-axis) time sequence during 1700-2018 period corresponding to 319-points (X-axis); Bottom: Contours of continuous wavelet coefficients colour-coded from minimum to maximum values with scales 1-128 (covering major FFT frequencies from Figure 4) for the same time period of 319-years.}
\end{figure}

It is known that the SSN peak year values vary for each solar cycle. The 11-year cycle is actually pseudo-periodic, and the past SSN data over centuries show that its amplitude of maxima varies by a factor of 2~to~3. The signal strength of the Gleissberg cycle (Frick et al., 1997) of $\approx$ 100 years duration is strong enough to modulate the main cycle of 11-years as can be seen from Figure 4. This modulation can be studied in more detail by subjecting the yearly mean SSN data to wavelet analysis. The FFT spectrum provides major frequency components with different sinusoidal wave periods but does not provide details of the relative magnitude of different prominent waves as a function of time. This is done by continuous wavelet analysis. The continuous wavelet transform is the sum over all time of the signal multiplied by scaled, shifted versions of the wavelet This process produces wavelet coefficients that are a function of scale and position (Torresani, 1997).  From the FFT analysis the main periods for the sunspot cycles are obtained as 11-, 22-, 53- and 107-years. The corresponding wavelet scales lie within 1-128 in a continuous one dimensional wavelet spectrum. 

Figure 5 shows this continuous wavelet spectrum of the yearly mean values of SSN time series during 1700-2018. It is clear from the figure that the maximum and minimum values of SSN related to different 11-year solar cycles go through a regular modulation of about 100-years. The wavelet spectrum shows very low values of coefficients for the entire range of scale/frequencies during three groups of solar cycles around 1800, 1900 and 2000 when the SSN peak values of corresponding solar cycles were about less than half of the normal peak SSN values. The effect of this coincidence for the first and second groups lasted for 2~to~3 solar cycles (approximately 30-years). From the pattern of variations of these coefficients, it is seen that the solar cycles 24 is going through this epoch of low solar activity. It is indicated that this third epoch may continue to the future solar cycles establishing the pattern of low solar activity trend of 2~to~3 solar cycles every 100-years. However this cannot be verified fully as existing predictive models at best project yearly mean sunspot numbers only for the next solar cycle 25 (Bhowmik et al., 2018; Hathway et al., 2016). An attempt has been made in our study to forecast sunspot numbers of solar cycle 25 running the publicly available HRNN model developed by Daniel Okoh and the results given in Okoh et al., 2018 mentioned earlier and generate a new set of continuous wavelet spectrum by extending the time series of annual mean SSN data to the year 2031 or to the end of solar cycle 25. 

\begin{figure}[h]
	\vspace*{0cm}
	\centering
	\makebox[0pt]{%
	\includegraphics[height=10cm,width=0.8\paperwidth]{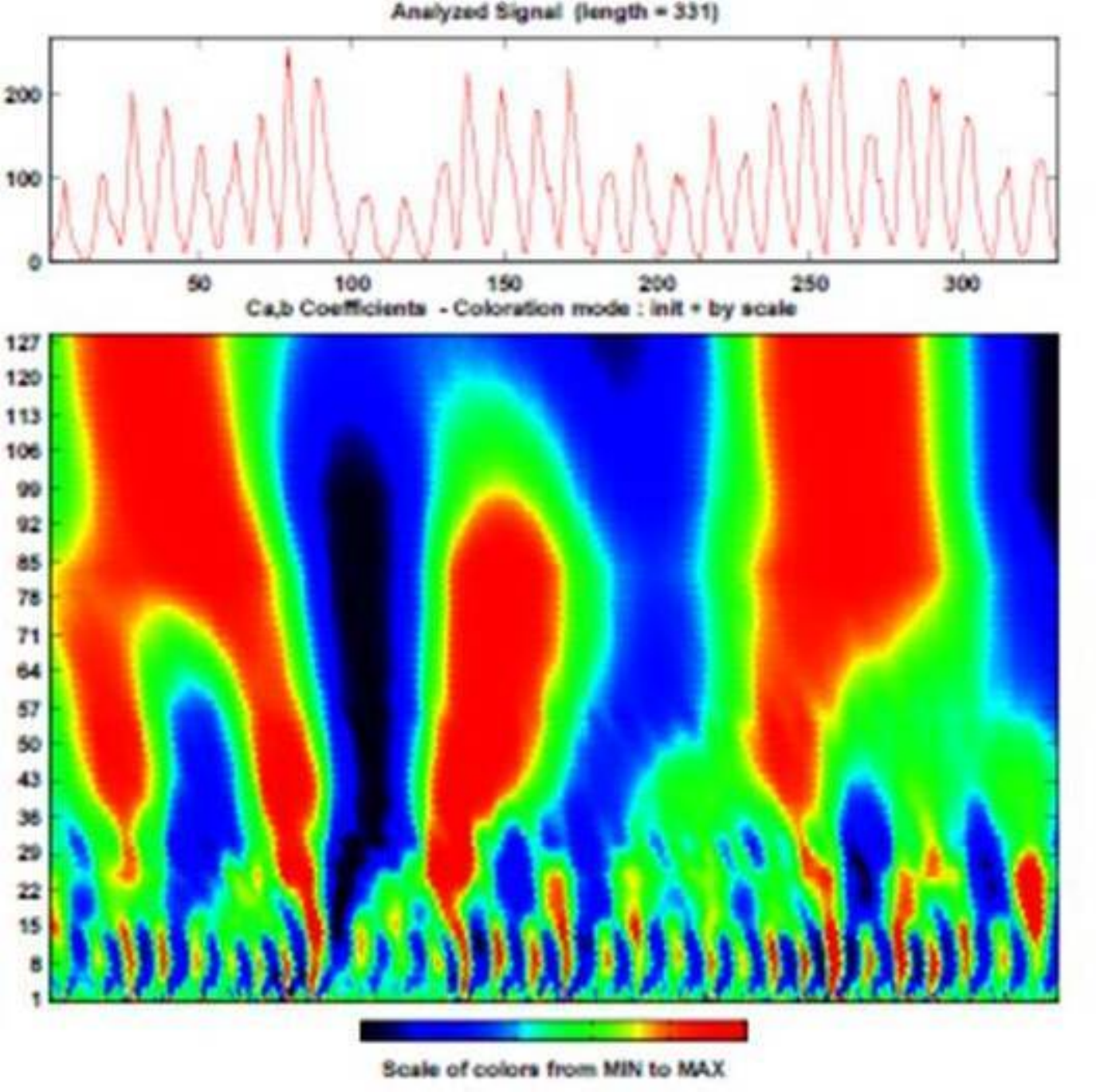}}
	\caption{Top: Yearly mean SSN (Y-axis) time sequence during 1700-2031 period corresponding to 331-points (X-axis); Bottom: Contours of continuous wavelet coefficients colour-coded from minimum to maximum values with scales 1-128 for the period 1700-2031, including the HRNN model generated annual mean SSN values between 2019-2031.}
\end{figure}

In HRNN model, regression analysis is combined with neural network learning for forecasting the SSN. The programme package consists of 2-executable files and one text file for inputs/outputs. Downloading this package from MathWorks, it is executed to obtain the smoothed yearly mean SSN values for the period 2019-2031 covering the solar cycle 25. The resultant predictions are added to the observed SSN series data till 2018 and the wavelet analysis is repeated for the entire range 1700-2031 (332-years). The result of such analysis is shown in Figure 6. It can be seen that the basic pattern of variation with the very low spell of solar activity is repeating roughly every 100-years. But the overall coincidence of low values of wavelet coefficients along the entire range of scales (vertical columns in Figure 5) continuing during the third group of low solar activity starting from solar cycle 24 indicates a higher probability of a situation like the first group of low solar activity. Hence from visual pattern recognition it is clear that the very low solar activity period is likely to continue not only up to 2031 but perhaps beyond during at least up to solar cycle 26 ($\approx$ 2031-2042), i.e., a spell of 3-solar cycles like the first group. While this is only a model prediction, it however implies that GCR radiation flux is likely to be near its peak at least during the solar cycle 25 and with little less certainty during the cycle 26 also. From the results of a predictive radiation model developed by Badhwar and O’Neill, 1992, it is estimated that the HZE particle fluxes inside a typical spacecraft in interplanetary space could be enhanced by a factor of 3 for the low SSN cycles like 24. Hence continuous and very high doses of deleterious cosmic radiation are expected at least up to 2032 taking into account the lag factor of one year shown earlier. It is known that while fluxes of HZE particles are only a fraction of the GCR small ions ($H$ and $He$), their LET or $dE/dx$ (a function of $Z^2$ where $Z$ being the ion charge or the atomic number) is very high (Chen et al., 1994) and coupled with their MeV-GeV energy range are very difficult to shield. Hence the current projections during solar cycle 25 provide a basis for estimating the cosmic radiation exposures in future long duration missions demonstrated in a statistical model of correlation between solar activity and GCR deceleration potential (Myung et al., 2007).

From the above results, it is clear that planning of any human Mars mission in near future should consider the long term cumulative effects of the high levels of GCR/HZE fluxes to minimise the exposure to maintain the acceptable career dose levels. Considerable research and technology development may be required to realise special shielding and other safeguarding strategies from these lethal radiations. As a follow up to this research work we intend to model the scenario of various shielding and other associated measures to carry out future exploration of Mars.

\section{Conclusion}
Future space missions to Mars would require long duration stay of at least 1 ~to~2 years by the astronauts for conducting any exploration and research. In such ventures one of the main challenges would be to limit the astronauts’ equivalent radiation doses to the NASA standard acceptable limits. However due to the unusual HZE fluxes characterised by very high energies, heavy ions and high LET, it is difficult to set a standard on the limiting values for biological effectiveness. Also it may be a very involved process to shield these cosmic radiations which are modulated by the solar activity. So the state of solar activity in the ensuing solar cycle 25 is predicted using a HRNN model. The yearly mean SSN data for 331-years (including the predicted values during 2019-2031) are subjected to Fourier and continuous 1-D wavelet analysis to predict the time sequence of the coefficients of various periodicities (e.g., 11-, 22-, 53- \& 107-years) and their possible effects on the peak values of 11-year cycle 25 and 26. It is found that apart from the main 11-year, Schwabe cycle, the SSN has a strong Gleissberg cycle of $\approx$ 100-years (here we see it centred around $\approx$ 107-years). The wavelet analysis has revealed that the years affected by continuously weak signals/coefficients of the entire range of scales/frequencies lead to a long duration spell of overall lower than normal SSN cycles such as cycles 24 and 25. Further research is being pursued to model spacecraft shielding, spacesuit and related safeguards for the astronauts should they venture a trip to Mars.

\section*{Acknowledgement}
The authors acknowledge the World Data Centre for Internationally revised monthly and annual mean sunspot numbers data products (ver~2) through Sunspot Index and Long-term Solar Observations (SILSO), Royal Observatory of Belgium, Brussels. The use of CRIS level-2 data sets from Advanced Composition Explorer (ACE) mission is also acknowledged. The data sets were archived and publicly available from ACE Science Center.

\section*{References}
\begin{description}
	
\item Babcock, H.W., The topology of the Sun’s magnetic field and the 22-year cycle, 1961, Astrophys. J., 133(2), 1961, 572-587.

\item Badhwar, G.D., O'Neill, P. M., An improved model of galactic cosmic radiation for space exploration missions, International Journal of Radiation Applications and Instrumentation; Part-D : Nuclear Tracks and Radiation Measurements, 20(3), 1992, 403-410.

\item Bertsch D.L., Fichtel, C.E., Reames, D. V., Relative abundance of iron-group nuclei in solar cosmic rays, Astrophys. J., 157, 1969, L53-L56.

\item Bhowmik, P., Nandy, D., Prediction of the strength and timing of sunspot cycle 25 reveals decadal-scale space environmental conditions, Nature Communications, 2018, doi: 10.1038/s41467-018-07690-0.

\item Borggräfe A., Quatmann, M., Nölke, D., Radiation protective structures on the base of a case study for a manned Mars mission, Acta Astronautica, 65, 2009, 1292-1305.

\item Chen, J., Chenette, d., Clark, R., Garcia-Munoz, M., Guzik, T. G., Pyle, K. R., Sang, Y., Wefel, J. P., A Model of galactic cosmic rays for use in calculating linear energy transfer spectra, Adv. Space Res., 14(10), 1994, 765-769.

\item Frick P., Galyagin, D., Foyt, D. V., Nesme-Ribes, E., Schatten, K. H., Sokoloff, D., Zakharov, V., Wavelet analysis of solar activity recorded by sunspot groups, Astron. \& Astrophys., 328, 1997, 670-681.

\item George, J. S., LaveK, A., Wiedenbeck, M. E., Binns, W. R., Cummings, A. C., Davis, A. J., de Nolfo, G. A., Hink, P. L., Israel, M. H., Mewaldt, R. A., Scott, L. M., Stone, E. C., von Rosenvinge, T.,Yanasak, N. E., Elemental composition and energy spectra of Galactic Cosmic Rays during cycle 23, Astrophys. J., 698(2), 2009, 1666-1681.

\item Hathaway, D.H., Lisa, A. U., Predicting the amplitude and hemispheric asymmetry of solar cycle 25 with surface flux transport, J. Geophys. Res. Space Physics, 121, 10, 2016, 744-753, doi:10.1002/2016JA023190

\item International Commission On Radiological Protection (ICRP), The Recommendations of the International Commission on Radiological Protection, 2007, 0146-6453.

\item Dietze, G., Bartlett, D. T., Cool, D. A., Cucinotta, F. A., Jia, X., McAulay, I. R., Pelliccioni, M., Petrov, V., Reitz, G., Sato, T., ICRP Publication 123: Assessment of Radiation Exposure of Astronauts in Space. Annals of the ICRP, 42(4), 1–339, 2013 doi: 10.1016/j.icrp.2013.05.004

\item Kim M.Y., Wilson, J. W., Cucinotta, F. A., A Solar cycle statistical model for the projection of space radiation environment, Adv. Space Res., 37(9), 2006, 1741-1748.

\item Lave, K. A., The Interstellar Transport of Galactic Cosmic Rays, Ph.D. Thesis, Washington University in St. Louis, 2012a, http://openscholarship.wustl.edu/etd/707.

\item Lave, K. A., Thesis Erratum, 2012b, http://www.srl.caltech.edu.

\item Kim, M. H. Y., de Angelis, G., Cucinotta, F. A., Probabilistic assessment of radiation risk for astronauts in space missions, Acta Astronautica, 68, 2011, 747–759.

\item Tony, C. S., Mertens, C. J., Blatting, S. R., Radiation Shielding Optimisation on Mars, NASA Langley Research Center, Hampton, Virginia, NASA/TP-2013-217985, 2013

\item O’Neill, P.M., Badhwar-O’Neill galactic cosmic ray model update based on advanced composition explorer (ACE) energy spectra from 1997 to present, Adv. Space Res., 37(9) 2006, 1727-1733, doi: 10.1016/j.asr.2005.02.001

\item Okoh, D. I., Seemala, G. K., Rabiu, A. B., Uwamahoro, J., Habarulema, J. B., Aggarwal, M., A Hybrid Regression‐Neural Network (HR‐NN) Method for Forecasting the Solar Activity, Space Weather, 16(9), 2018, doi: 10.1029/2018SW001907

\item Simpson, J. A., Elemental and isotopic composition of the galactic cosmic rays, Ann. Rev. Nucl. Part. Sci., 33, 1983, 323–381.

\item Singleterry, R. C., Blattnig, S. R., Clowdsley, M. S., Qualls, G. D., Sandridge, C. A., Simonsen, L. C., Norbury, J. W., Slaba, T. C., Walker, S. A., Badavi, F. F., Spangler, J. L., Aumann, A. R., Zapp, E. N., Rutledge, R. D., Lee, K. T., Norman, 
R. B., OLTARIS: On-Line Tool for the Assessment of Radiation in Space, NASA/TP–2010-216722, 2010

\item Stone, E. C., Frandsen, A. M., Mewaldt, R. A., Christian, E. R., Margolies, D., Ormes, J. F., Snow, F., The Advanced Composition Explorer, Space Sci. Rev., 86, 1998, 1-22. 

\item Torresani, B., Analyse continue par ondelettes, Inter-editions, Savior Actuels, 1995. 
\end{description}

\end{document}